\documentclass[final,1p,times,sort&compress]{elsarticle} 
\usepackage{graphicx}
\usepackage{amssymb} 
\usepackage{amsthm} 
\usepackage{color}
\definecolor{darkgreen}{rgb}{0,.7,0}
\definecolor{linkblue}{rgb}{0.,0.,0.9333}
\usepackage{hyperref} %create internal and external hyperlinks
\hypersetup{
    pdffitwindow=true,      % page fit to window when opened
    pdfauthor={W. A. Horowitz},     % author
    pdfnewwindow=true,      % links to different pdfs in new window
    bookmarksnumbered=true,
    colorlinks=true,       % false: boxed links; true: colored links
    linkcolor=red,          % color of internal links
    citecolor=darkgreen,        % color of links to bibliography
    filecolor=magenta,      % color of file links
    urlcolor=cyan,           % color of external links
    menucolor=blue,
}

\newcommand{\fig}[1]{Fig.\ \ref{#1}}
\newcommand{\eq}[1]{Eq.\ (\ref{#1})}

\newcommand{\tauform}{\tau_{form}}

\journal{Nuclear Physics A} 

\begin{document}

\begin{frontmatter} 

% Your Title - please insert
\title{Heavy Quark Production and Energy Loss}

%% Single author (and collaboration) - please insert
\author{W.\ A.\ Horowitz}
\address{Department of Physics, University of Cape Town, Private Bag X3, Rondebosch 7701, South Africa}
\ead{wa.horowitz@uct.ac.za}
%\ead[url]{http://www.phy.uct.ac.za/people/horowitz}

\begin{abstract} 
Heavy flavor research is a vigorous and active topic in high-energy QCD physics.  Comparing theoretical predictions to data as a function of flavor provides a unique opportunity to tease out properties of quark-gluon plasma.  We explicitly demonstrate this utility with energy loss predictions based on the assumption of 1) a weakly-coupled plasma weakly coupled to a high-$p_T$ probe using pQCD and 2) a strongly-coupled plasma strongly coupled to a high-$p_T$ probe using AdS/CFT; we find that while the former enjoys broad qualitative agreement with data, it is difficult to reconcile the latter with experimental measurements.
\end{abstract} 

\end{frontmatter} % do not change

%% linenumbers are useful for reviewing process
%\linenumbers

\section{Introduction}
Our goal as nuclear physicists is to quantitatively extract experimentally and understand theoretically the properties of nuclear matter; as part of this goal, we wish to compose the phase diagram of the strong force.  This is an extremely immodest goal.  For example, detailed first-principles calculations for the phase diagram of hydrogen, the most simple QED system, is contemporary research \cite{Militzer}.  Nevertheless quark-gluon plasma (QGP) provides us with a unique opportunity to probe experimentally and theoretically the emergent, many-body physics of a non-Abelian gauge theory in a certain region of its phase diagram.  We have a number of tools at our disposal for exploring the properties of QGP from experimental measurements, for example: low-$p_T$ particles, electromagnetic probes, quarkonia, high-$p_T$ light hadrons, and high-$p_T$ heavy hadrons.  
Heavy flavor measurements and theory are an important piece in the puzzle we are trying to put together to form a consistent and coherent picture of heavy ion collision phenomena.

Unfortunately length limits the breadth of the coverage here of the fascinating and important aspects of heavy flavor research in high-energy QCD; we will focus on the physics we can learn from high momentum particles.  High-$p_T$ particles are especially interesting as they are the decay products of high-$p_T$ partons, which are the most direct probe of the relevant degrees of freedom in a quark-gluon plasma (QGP) \cite{Majumder:2010qh,CasalderreySolana:2011us}.  
One is able in principle to learn about QGP by making an assumption regarding the physics of the QGP and comparing the necessary theoretical consequences of those assumptions to data.  One hopes to use this approach to falsify certain assumed descriptions of the plasma and add evidence for others.  Just as one requires consistency of the entire picture of all data, one requires consistency of description of measurements associated with energy loss.  It turns out that this consistency is quite hard to achieve, and, as a result, energy loss provides us with a valuable window through which to actually measure the physics of quark-gluon plasma.  To explicitly demonstrate the power of this method we will take two extreme, generic assumptions regarding the medium and how it couples to high-$p_T$ probes and compare the results to data. 

Heavy flavor is especially interesting because of the additional experimental constraints it puts on energy loss calculations, and thus on the potential properties of quark-gluon plasma.  In particular, an energy loss calculation, in a broad sense, results in a probability of a parton losing some of its initial momentum, $P(\Delta p_T\,|\,p_T,\,L,\,T,\,M_Q,\,R)$, where $L$ is the pathlength the parent parton travels, $T$ is the temperature of the plasma, $M_Q$ is the mass (or effective mass) or the parent parton, and $R$ is the representation (i.e.\ is the parent parton a gluon or quark).  Unfortunately one cannot alter these parameters experimentally to test energy loss theories; rather one changes, for instance, the collision species, the $\surd s$ of the collision, or---as is especially useful for investigating heavy flavor---the mass of the measured hadron, and compares to theoretical predictions.

\begin{figure}[!htbp]
	\begin{center}
		$\begin{array}{cc}
			\includegraphics[width=0.49\textwidth]{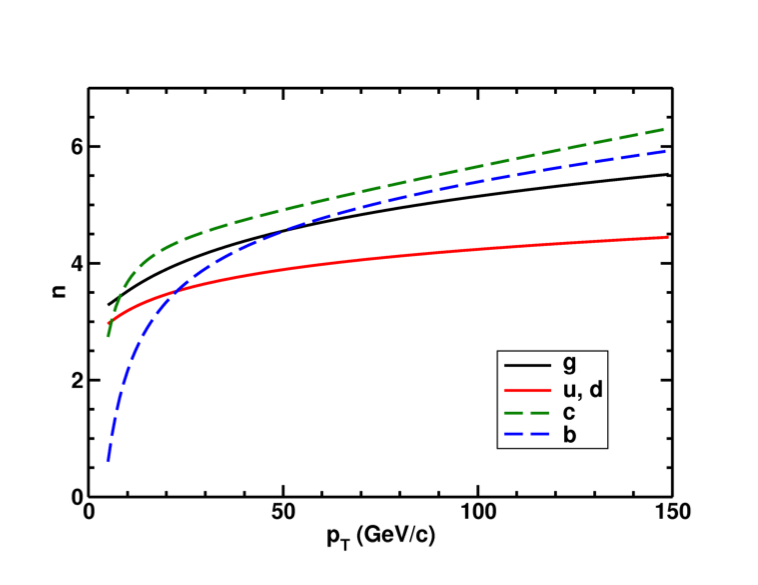} & \includegraphics[width=0.49\textwidth]{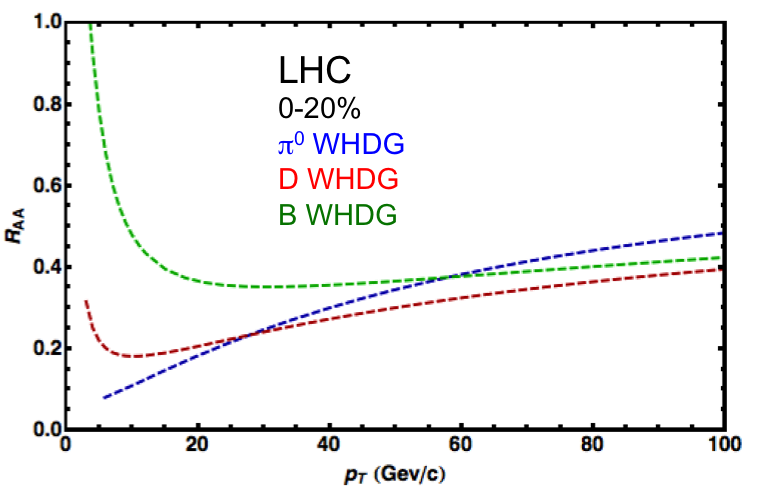} \end{array}$
	\end{center}
	\vspace{-.3in}$\hspace{.18in} \mbox{(a)} \hspace{2.5in} \mbox{(b)}$
	\caption{(a) The power $n_Q(p_T)+1$ that best approximates the production spectra for various partonic species $Q$ \cite{Wicks:2005gt} and (b) $R_{AA}(p_T)$ for pions, charm mesons, and bottom mesons at 0-20\% centrality collisions \cite{Horowitz:2011cv}.  Both plots use $\surd s_{NN}=2.76$ ATeV.}
	\label{f:nandcross}
\end{figure}

Qualitatively, one expects a simple ordering of the energy losses as one changes parent parton species, specifically\footnote{It's worth noting for the experts that this ordering is not strict in pQCD: because the formation time has mass dependence \cite{Djordjevic:2003zk} there can be violations in the ordering in certain regions of phase space.}
\begin{equation}
	\Delta E_b < \Delta E_c < \Delta E_{u,\,d,\,s} < \Delta E_g.
\end{equation}
However, it is important to emphasize that \emph{energy loss ordering does not necessitate an ordering of the nuclear suppression factors}, $R_{AA}(p_T)$, because $R_{AA}$ involves a convolution over the fragmentation functions and, most important, the production spectrum.  For approximately power law production of high-$p_T$ partons, as predicted by pQCD, and fractional momentum loss $\epsilon = (p_{T,\,i}-p_{T,\,f})/p_{T,\,i}$, one can show that \cite{Horowitz:2010dm}
\begin{equation}
	\label{e:raa}
	\frac{dN}{dp_T}\propto \frac{1}{p_T^{n+1}} \quad \Rightarrow \quad R_{AA}\approx\big\langle\int d\epsilon(1-\epsilon)^n P(\epsilon)\big\rangle,
\end{equation}
where the angular brackets refer to averaging over the geometry.  As one can see from \eq{e:raa}, the softer the spectra the larger is $n$ and thus the smaller $R_{AA}$ for the same $\Delta p_T$.  As seen in \fig{f:nandcross}, the relative changes in production power laws $n_Q$ for different parton species lead to a crossing in $R_{AA}$ for different hadron species (as first noted in \cite{Buzzatti:2011vt}).

The use of heavy flavor allows for an additional test of energy loss theory, and it turns out that by comparing to the centrality, center of mass energy, and especially flavor dependence of data yields a stringent constraint on the theory; see how, for instance, a simultaneous description of pion and non-photonic electron suppression at RHIC precludes the possibility of pQCD-based radiative only energy loss \cite{Djordjevic:2005db}.
\section{Extracting Physics by Comparing to Data}
\subsection{Strongly-coupled Medium Strongly Coupled to a Probe}
To begin our energy loss comparison to data using two extreme assumptions about the physics of the QGP, let's consider a strongly-coupled medium strongly coupled to the high-$p_T$ probe.  There are many reasons to believe that strong-coupling dynamics dominant the physics of the medium, and in particular, that AdS/CFT techniques provide valuable insight into these processes \cite{Gubser:2009fc, CasalderreySolana:2011us}.  For instance, running coupling calculations suggest that at $T\sim250$ MeV---a not unreasonable placeholder for the QGP temperature---$g\sim2$ and $\lambda=g^2N_c\sim12\gg1$; it's worth noting that in phenomenological applications $T$ is never large compared to $\Lambda_{QCD}$.  Also, for $T\gtrsim T_c$, lattice calculations nontrivially deviate from the Stefan-Boltzmann limit, and in such a way that is reasonably well described using AdS/CFT \cite{Gubser:2009fc}.  Finally, the viscosity to entropy ratio extracted from hydrodynamics calculations \cite{Heinz:2011kt} suggest $\eta/s\sim1/4\pi$, which is readily explained by AdS/CFT \cite{Kovtun:2004de}.  

\begin{figure}[!htbp]
	\begin{center}
		$\begin{array}{ccc}
			\includegraphics[width=.45\textwidth]{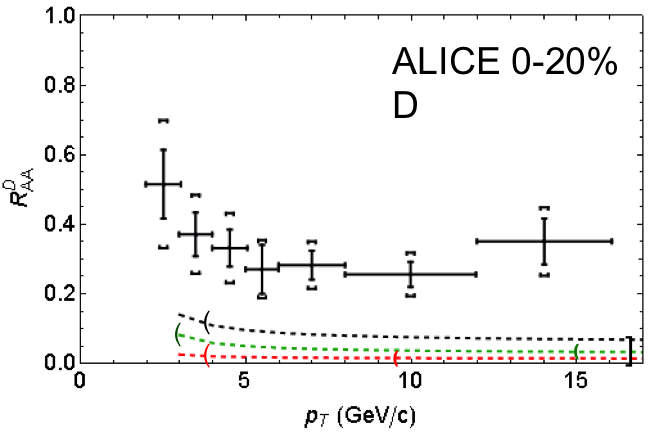} & & \includegraphics[width=.45\textwidth]{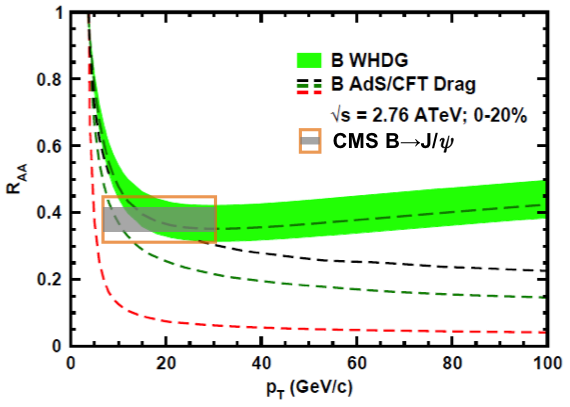}
		\end{array}$
	\end{center}
	\vspace{-.2in}$\hspace{.2in} \mbox{(a)} \hspace{2.8in} \mbox{(b)}$
	\caption{Comparison of (a) $D$ \cite{ALICE:2012ab} and (b) $B$ \cite{Chatrchyan:2012np} meson $R_{AA}(p_T)$ for 0-20\% centrality collisions at LHC to AdS/CFT heavy quark drag predictions constrained by RHIC data \cite{Horowitz:2007su,Horowitz:2010dm,Horowitz:2011wm}.}
	\label{f:AdSheavies}
\end{figure}

There have been calculations of the energy loss of both light and heavy quarks using the AdS/CFT correspondence.  For heavy flavor, the energy loss is a drag, $dp_T/dt=-\mu p_T$, where $\mu =\pi\,\lambda^{1/2}T^2/2M_Q$ \cite{Herzog:2006gh,Gubser:2006bz}; this is similar to weakly-coupled energy loss in the Bethe-Heitler regime, but very different from the predictions of pQCD in the deep LPM region where $dp_T/dt\sim-L\,T^3\ln(p_T/M_Q)$ \cite{Gyulassy:2000er}.  Comparison between AdS/CFT calculations and data are difficult because there is no unique mapping from the parameters of QCD to those of $\mathcal{N}=4$ SYM and AdS$_5\times S^5$.  Nevertheless, comparing over reasonable assumptions for parameter values in AdS/ CFT yields a quantitative agreement between theoretical predictions and data for non-photonic electron suppression at RHIC \cite{Akamatsu:2008ge,Horowitz:2007su,Horowitz:2010dm}.  But we see again the power of comparing theoretical calculations to a wide range of data when we attempt to simultaneously describe the suppression of heavy flavor at LHC.  Keeping all parameters fixed and only changing the temperature of the medium (which we do according to the measured $\surd s$ dependence of the multiplicity) we can calculate zero parameter predictions for LHC, shown in \fig{f:AdSheavies}.  While the $B$ meson suppression is currently consistent with data within the large experimental and theoretical uncertainties, the AdS/CFT calculations significantly overpredict the suppression of $D$ mesons.  Before strong conclusions are drawn, however, it turns out that momentum fluctuations \cite{Gubser:2006nz,CasalderreySolana:2007qw} (especially longitudinal)---whose importance should only affect momenta parametrically large compared to the momenta at which the formalism breaks down, and were neglected in these calculations---likely play a significant role numerically.

One wants not only to simultaneously compare AdS/CFT predictions to data as a function of $\surd s$ but also as a function of parton species.  Unfortunately the theory of light flavor energy loss \cite{Gubser:2008as,Chesler:2008uy,Ficnar:2012nu} is less well understood in AdS/CFT than for heavy quarks.  Added difficulties arise in the light sector due to the lack of 1) an analytic solution for falling string configurations and 2) a good working definition for the energy lost by the probe (in principle one can exactly compute $T^{\mu\nu}$ for the plasma and thus the energy lost by the probe, but this is an extremely difficult problem both in terms of the analytics and the numerics).  Preliminary estimates, however, suggest that light flavor energy loss is also overpredicted by AdS/CFT: the thermalization time for light quarks in the medium is of the order of 3 fm; even when the 1D Hubble flow of the QGP is included the thermalization time is only increased to about 4 fm \cite{MoradHorowitz}.  

In addition to checking the flavor dependence, these AdS/CFT energy loss calculations may also be tested by changing collision species and looking at the suppression of heavy flavor at forward rapidities \cite{Horowitz:2009pw} (at mid-rapidity, open heavy flavor production can also test saturation effects \cite{Dominguez:2011cy}).  Other information on the use of AdS/CFT in heavy ion collisions discussed during the conference can be found at \cite{Chesler:2010bi,Ficnar:2012nu,Chesler:2011nc}.

\subsection{Weakly-coupled Medium Weakly Coupled to a Probe}
The assumption of the dominance of weakly-coupled dynamics in heavy ion collisions is also not unreasonable.  For $T\sim250$ MeV, $\alpha_s(2\pi T) = 0.3$.  Also, multi-loop thermal field theory is in good agreement with lattice data for thermodynamic properties of QCD at a few times $T_c$, albeit with large uncertainties \cite{Andersen:2011sf}.  Finally, numerically daunting parton cascade calculations show that including $2\rightarrow3$ channels yields $\eta/s\sim$ few$/4\pi$ \cite{El:2008yy}.

Continuing with the assumption of a weakly-coupled plasma coupling weakly to a probe, the medium is described by two scales: the Debye screening length, given in terms of the Debye mass $\mu\sim gT$, and a mean free path for gluons, $\lambda_{mfp}^g\sim 1/g^2T$ (see \cite{Wicks:2005gt} and references therein).  When evaluated at temperature scales relevant for RHIC and LHC and with all the numerical coefficients, one finds\footnote{For $T_{RHIC}=350$ MeV ($T_{LHC}=450$ MeV), $1/\mu\simeq0.3\,(0.2)$ fm and $\lambda_{mfp}^g\simeq0.8\,(0.7)$ fm.} an ordering of scales $1/\mu\ll\lambda_{mfp}^g\ll L$, where again $L$ is the pathlength travelled by the parent parton and is on the scale of the radius of the nucleus, $L\sim R_{A}$.  Since $1/\mu\ll\lambda_{mfp}^g$, high-$p_T$ particles scatter off of well defined, separated medium quasi-particles.    It is important to note however that for heavy quarks $L/\lambda_{mfp}\sim4$ (and similarly even for gluons), and therefore energy loss models that assume a large number of collisions (and that thus the central limit theorem holds), such as those using Langevin or rates methods, likely require large corrections.

\begin{figure}[!htbp]
	\begin{center}
		$\begin{array}{ccc}
			\includegraphics[width=.45\textwidth]{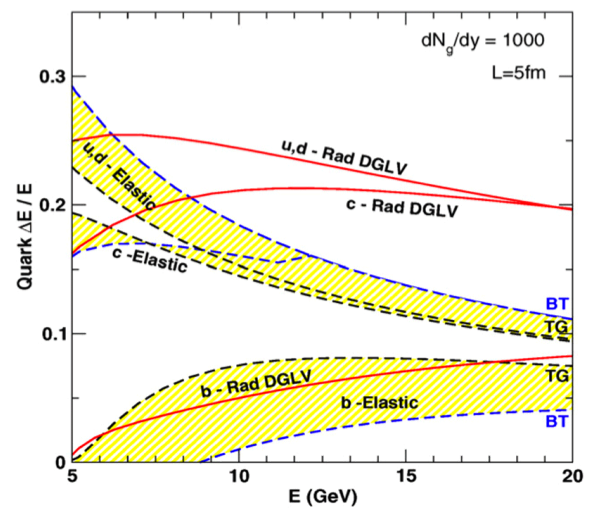} & & \includegraphics[width=.45\textwidth]{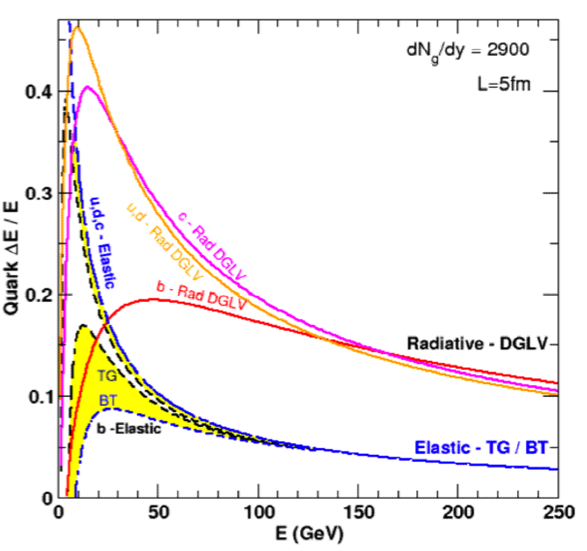}
		\end{array}$
	\end{center}
	\vspace{-.2in}$\hspace{.2in} \mbox{(a)} \hspace{2.8in} \mbox{(b)}$
	\caption{Elastic and radiative energy loss for gluons and light, charm, and bottom quarks travelling a distance $L=5$ fm through QGP at (a) RHIC and (b) LHC temperatures \cite{Horowitz:2010dm}.}
	\label{f:WHDGeloss}
\end{figure}

In pQCD with its quasi-particle picture one can distinguish between two types of energy loss: elastic and inelastic, otherwise known as collisional and radiative, respectively.  There is a long history of pQCD-based elastic energy loss calculations (see \cite{Horowitz:2010dm} and references therein); see \cite{Majumder:2010qh} for a review of pQCD-based radiative energy loss calculations.  Leading order estimates of the size of elastic energy loss yield $d p_T^{el}/dt\sim -T^2\ln(p_T/M_Q)$.  Naively, at asymptotically large energies intuition based on classical electromagnetism leads to the conclusion that $\Delta E_{el}\ll\Delta E_{rad}$, but this is based on a Bethe-Heitler estimate of radiative energy loss in which subsequent collisions with medium particles yield incoherently summable emissions.  However, there is one more important scale to qualitatively understand radiative energy loss, the formation time, $\tau_{form}$, which characterizes the distance required for an emitted gluon to be resolved independently from the emitting parton.  There is a large uncertainty in the size of $\tauform$ but for emissions of large energy gluons in QGP, $\tauform\gg\lambda_{mfp}^g$, 
and a single gluon emission is thus produced from coherent scatterings off of multiple in-medium quasi-particles. 
This reduction in the amount of emitted radiation is known as the Landau-Pomeranchuk-Migdal, or LPM, effect \cite{Majumder:2010qh}.  In the LPM limit $\Delta E_{rad}\propto-L\,T^3\ln(p_T/M_Q)$.  With this reduction in radiative energy loss it is possible for elastic energy loss to be important even at asymptotically high energies.  We will be using the WHDG model of convolved radiative and elastic energy loss \cite{Wicks:2005gt} for explicit comparison to experimental data; in this calculation which uses thermal field theory methods for computing the elastic energy loss and the DGLV derivation for the radiative, the elastic energy loss remains a significant contributor to total energy loss even for 250 GeV/c partons at LHC; see \fig{f:WHDGeloss}.

\begin{figure}[!htbp]
	\begin{center}
		\includegraphics[width=\textwidth]{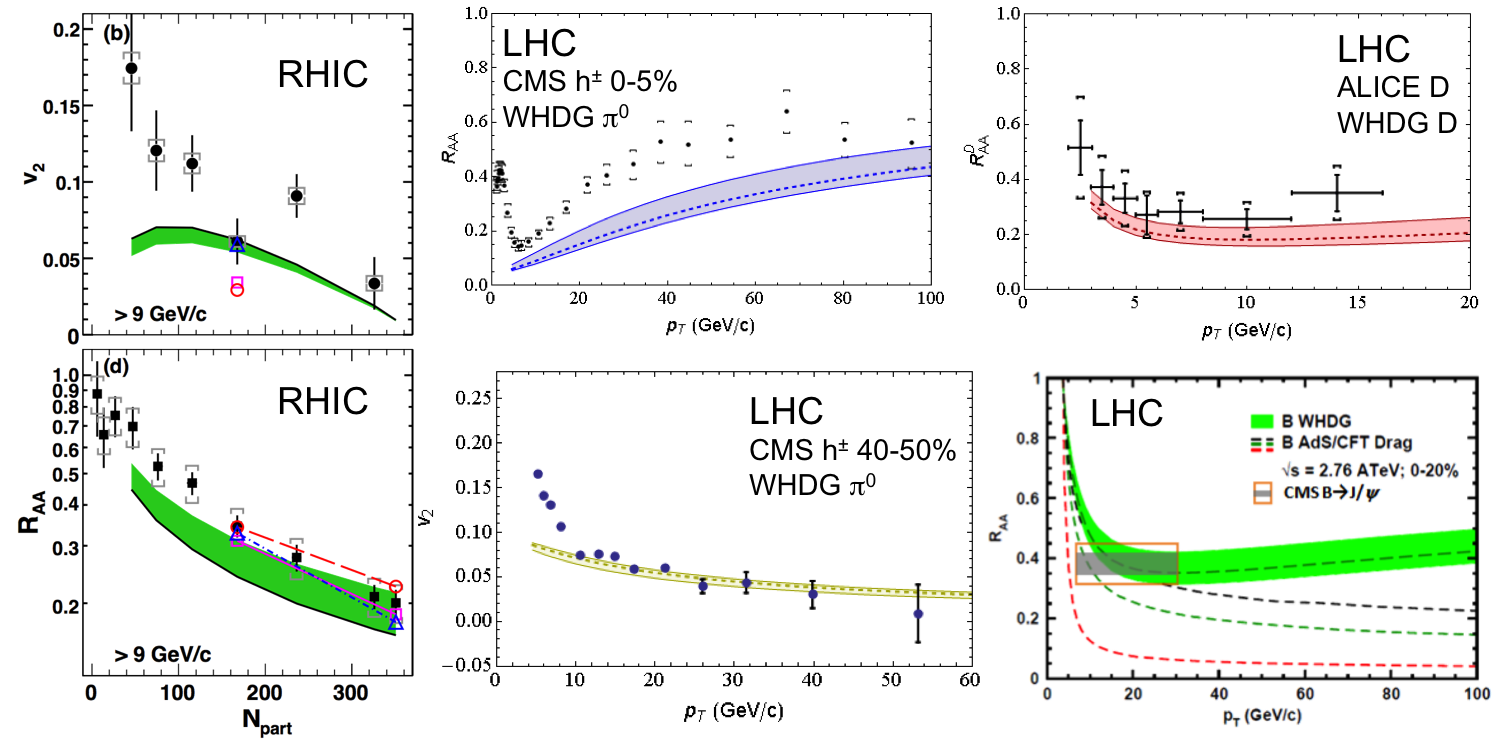}
	\end{center}
	\vspace{-1.7in}
	$\hspace{.06in} \mbox{(a)} \hspace{1.3in} \mbox{(b)} \hspace{1.8in} \mbox{(c)}$\\[1.2in]
	$\mbox{\phantom{b}(d)} \hspace{1.3in} \mbox{(e)} \hspace{1.88in} \mbox{(f)}$
	\caption{Constrained zero parameter WHDG predictions compared to data for (a) $v_2(N_{part})$ at RHIC \cite{Adare:2010sp,Wicks:2005gt}, (b) 0-5\% centrality $R_{AA}(p_T)$ for light flavors at LHC \cite{CMS:2012aa,Horowitz:2011gd}, (c) $R_{AA}^D(p_T)$ at 0-20\% centrality at LHC \cite{ALICE:2012ab,Horowitz:2011cv}, (d) $R_{AA}^{\pi^0}(N_{part})$ at RHIC \cite{Adare:2010sp,Wicks:2005gt}, (e) $v_2(p_T)$ at LHC for light flavors at 40-50\% centrality \cite{Chatrchyan:2012xq,Horowitz:2011cv}, and (f) $R_{AA}^B(p_T)$ at 0-20\% centrality at LHC \cite{Chatrchyan:2012np,Horowitz:2011wm}.}
	\label{f:WHDGandData}
\end{figure}

We would like to compare the WHDG model to as many observables as possible.  Using the thermal field theoretic methods to relate $\mu$ and $\lambda_{mfp}^g$ to temperature, assuming that the temperature profile is proportional to the Glauber participant density, and that all couplings are approximately fixed at $\alpha_s=0.3$ there is only one free parameter in the theory, the proportionality constant relating the observed multiplicity to the entropy of the plasma.  The PHENIX experiment rigorously extracted the best fit value of this parameter and its uncertainty, the rapidity density of gluons $dN_g/dy = 1400_{-375}^{200}$ \cite{Adare:2008cg}, by comparing to their $R_{AA}^{\pi^0}(p_T)$ measurement in most central $\surd s = 200$ AGeV collisions.  Before immediately comparing to the multitude of data from RHIC and LHC, it is worth noting that the lack of precision and accuracy inherent in pQCD due to both the complicated nature of the theory and the relatively large size of its coupling constant: even NLO calculations of production rates in hadronic collisions tend to be correct only within a factor of 2 of the data \cite{Adare:2010de,CMS:2012aa}.  With this in mind, the agreement between the LO WHDG energy loss theory and data shown in \fig{f:WHDGandData}
over a range of centralities, collision energies, measurements, and flavors is surprisingly good.

There are a number of directions in which these perturbative calculations can be improved, many of which were discussed at the conference.  For instance one might try to model the energy loss using a parton cascade, which trades a better treatment of multiple gluon emission for a less accurate treatment of the quantum mechanical formation time effects \cite{Uphoff:2012it}. Or one might attempt a NLO ansatz for running coupling along with a more careful treatment of production spectra and time evolution \cite{Buzzatti:2012pe}.  For heavy flavors, some authors have attempted to include additional energy loss channels such as in-medium fragmentation \cite{Sharma:2012dy} or non-perturbative cross sections \cite{He:2012xz}.  Other heavy flavor calculations shown at this conference include \cite{Cao:2012as,Gossiaux:2012cv}.  However, one should take care with calculations that include uncontrolled physics, especially ones that are in principle uncontrollable; it is not clear what information can be learned when unconstrained processes are involved.  These include calculations in which parameters are dialled in to values that are not limited by data or are inconsistent with the assumptions underlying the model or when important energy loss channels are neglected (i.e.\ an energy loss model includes only radiative or elastic loss).  It's also worth emphasizing that assuming a perturbative picture necessarily implies that the in-medium induced radiative energy loss is a result of the interference between the collinearly and infrared divergent radiation associated with the production of a bare color charge and the radiation that is created by stimulated emission from collisions of this bare color charge with colored objects in the QGP; calculations that do not incorporate this interference physics necessarily have the wrong energy and length dependencies.

\subsection{Direct Comparison of the Pictures}
\begin{figure}
	\begin{center}
		\includegraphics[width=.6\textwidth]{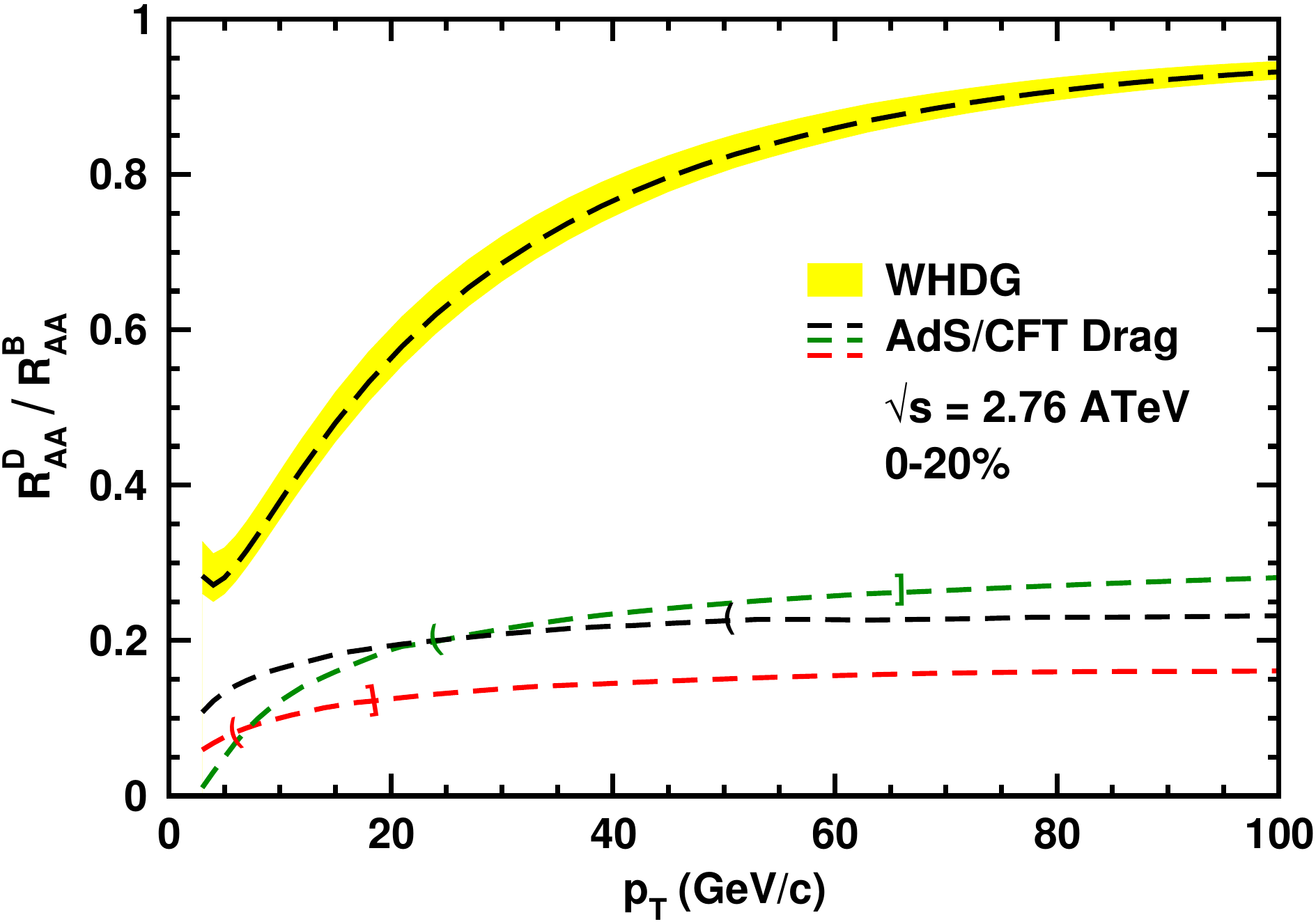}
	\end{center}
	\vspace{-.1in}
	\caption{Comparison of the double ratio of $D$ meson to $B$ meson $R_{AA}(p_T)$ for 0-20\% centrality collisions at LHC using the pQCD-based WHDG energy loss model \cite{Wicks:2005gt,Horowitz:2011gd} and a model based on AdS/CFT drag energy loss \cite{Horowitz:2007su,Horowitz:2011wm}.}
	\label{f:dtob}
\end{figure}

Although the high-$p_T$ physics evidence for a weakly-coupled plasma weakly coupled to a probe is strong and there are significant signs of disagreement between the predictions of a strongly-coupled plasma strongly coupled to a probe it is worth considering a measurement that shows a qualitative difference between the two pictures.  One may emphasize the different mass and momentum dependencies of the pQCD and AdS/CFT results by considering the double ratio of $D$ to $B$ meson $R_{AA}$ as seen in \fig{f:dtob}.  While the leading order AdS/CFT results are applicable (up to a speed limits indicated on the graph) the mass dependence of the energy loss remains; on the other hand the mass dependence drops out for the perturbative results at asymptotically large momenta.

\section{Summary}
We seek a coherent, consistent picture of the physics of QGP in our quest to understand its properties.  The comparison of energy loss calculations to data provide a direct probe of the relevant degrees of freedom in a QGP and how this physics interacts with high-$p_T$ particles.The flavor dependence is especially useful for discriminating between QGP assumptions as their predictions vary significantly as a function of high-$p_T$ parton masses.  As demonstrated by the participants at this conference, heavy flavor phenomenology is rich, varied, and vigorous.

As we have known since the first anisotropy and heavy flavor results from RHIC, a simultaneous description of multiple observables related to energy loss physics is very hard to achieve.  In particular, despite successes at RHIC, predictions for LHC based on the strong coupling physics of AdS/CFT do not appear to describe the data, although possibly important physics was neglected.  However, LO pQCD results give a rather good qualitative description of a suite of observables including pion and heavy flavor suppression and anisotropy from RHIC to LHC.  Should we find that these tentative conclusions hold, it becomes a very interesting question of how the strong-coupling low-$p_T$ physics of the QGP medium as implied by hydrodynamics comparisons to data turn over to weak-coupling physics of the QGP medium as implied by the energy loss calculations.

\section{Acknowledgments}
Support from the National Research Foundation of South Africa and SA-CERN is gratefully acknowledged.


\section*{References}

\begin{thebibliography}{10}
\expandafter\ifx\csname url\endcsname\relax
  \def\url#1{\texttt{#1}}\fi
\expandafter\ifx\csname urlprefix\endcsname\relax\def\urlprefix{URL }\fi
\expandafter\ifx\csname href\endcsname\relax
  \def\href#1#2{#2} \def\path#1{#1}\fi

\bibitem{Militzer}
B.~Militzer, Path integral monte carlo simulations of hot dense hydrogen, Ph.D.\
  thesis, University of Illinois, Urbana-Champaign (2000).

\bibitem{Majumder:2010qh}
A.~Majumder, M.~Van~Leeuwen, Prog.Part.Nucl.Phys.\ A66 (2011) 41--92, \href{http://arxiv.org/abs/1002.2206}{arXiv:1002.2206}.
%%CITATION = ARXIV:1002.2206;%%

\bibitem{CasalderreySolana:2011us}
J.~Casalderrey-Solana, H.~Liu, D.~Mateos, K.~Rajagopal, U.~A. Wiedemann, \href{http://arxiv.org/abs/1101.0618}{arXiv:1101.0618}.
%%CITATION = ARXIV:1101.0618;%%

\bibitem{Djordjevic:2003zk}
M.~Djordjevic, M.~Gyulassy, Nucl.\ Phys.\ A733 (2004) 265--298, \href{http://arxiv.org/abs/nucl-th/0310076}{arXiv:nucl-th/0310076}.
%%CITATION = NUCL-TH/0310076;%%

\bibitem{Horowitz:2010dm}
W.~A. Horowitz, Probing the Frontiers of QCD, Ph.D.\ thesis, Columbia University (2008), \href{http://arxiv.org/abs/1011.4316}{arXiv:1011.4316}.
%%CITATION = ARXIV:1011.4316;%%

\bibitem{Buzzatti:2011vt}
A.~Buzzatti, M.~Gyulassy, Phys.Rev.Lett.\ 108 (2012) 022301, \href{http://arxiv.org/abs/1106.3061}{arXiv:1106.3061}.
%%CITATION = ARXIV:1106.3061;%%

\bibitem{Djordjevic:2005db}
M.~Djordjevic, M.~Gyulassy, R.~Vogt, S.~Wicks, Phys.Lett.\ B632 (2006)
  81--86, \href{http://arxiv.org/abs/nucl-th/0507019}{arXiv:nucl-th/0507019}.
%%CITATION = NUCL-TH/0507019;%%

\bibitem{Gubser:2009fc}
S.~S. Gubser, Nucl.Phys.\ A830 (2009) 657C--664C, \href{http://arxiv.org/abs/0907.4808}{arXiv:0907.4808}.
%%CITATION = ARXIV:0907.4808;%%

\bibitem{Heinz:2011kt}
U.~Heinz, C.~Shen, H.-C. Song, AIP Conf.Proc.\ 1441 (2012) 766--770, \href{http://arxiv.org/abs/1108.5323}{arXiv:1108.5323}.
%%CITATION = ARXIV:1108.5323;%%

\bibitem{Kovtun:2004de}
P.~Kovtun, D.~Son, A.~Starinets, Phys.Rev.Lett.\ 94 (2005) 111601, \href{http://arxiv.org/abs/hep-th/0405231}{arXiv:hep-th/0405231}.
%%CITATION = HEP-TH/0405231;%%

\bibitem{Herzog:2006gh}
C.~Herzog, A.~Karch, P.~Kovtun, C.~Kozcaz, L.~Yaffe, JHEP 0607 (2006)
  013, \href {http://arxiv.org/abs/hep-th/0605158}
  {arXiv:hep-th/0605158}.
%%CITATION = HEP-TH/0605158;%%

\bibitem{Gubser:2006bz}
S.~S. Gubser, Phys.Rev.\ D74 (2006) 126005, \href {http://arxiv.org/abs/hep-th/0605182}
  {arXiv:hep-th/0605182}.
%%CITATION = HEP-TH/0605182;%%

\bibitem{Gyulassy:2000er}
M.~Gyulassy, P.~Levai, I.~Vitev, Nucl.Phys.\ B594 (2001) 371--419,
\newblock \href {http://arxiv.org/abs/nucl-th/0006010}
  {arXiv:nucl-th/0006010}.
%%CITATION = NUCL-TH/0006010;%%

\bibitem{Akamatsu:2008ge}
Y.~Akamatsu, T.~Hatsuda, T.~Hirano, Phys.Rev.\ C79 (2009) 054907,
\newblock \href {http://arxiv.org/abs/0809.1499} {arXiv:0809.1499}.
%%CITATION = 0809.1499;%%

\bibitem{Horowitz:2007su}
W.~Horowitz, M.~Gyulassy, Phys.Lett.\ B666 (2008) 320--323,
\newblock \href {http://arxiv.org/abs/nucl-th/0706.2336}
  {arXiv:nucl-th/0706.2336}.
%%CITATION = 0706.2336;%%

\bibitem{Gubser:2006nz}
S.~S. Gubser, Nucl.Phys.\ B790 (2008) 175--199,
\newblock \href {http://arxiv.org/abs/hep-th/0612143}
  {arXiv:hep-th/0612143}.
%%CITATION = HEP-TH/0612143;%%

\bibitem{CasalderreySolana:2007qw}
J.~Casalderrey-Solana, D.~Teaney, JHEP 0704 (2007) 039,
\newblock \href {http://arxiv.org/abs/hep-th/0701123}
  {arXiv:hep-th/0701123}.
%%CITATION = HEP-TH/0701123;%%

\bibitem{ALICE:2012ab}
B.~Abelev, et~al., JHEP 1209 (2012) 112,
\newblock \href {http://arxiv.org/abs/1203.2160} {arXiv:1203.2160}.
%%CITATION = 1203.2160;%%

\bibitem{Chatrchyan:2012np}
S.~Chatrchyan, et~al., JHEP 1205 (2012) 063,
\newblock \href {http://arxiv.org/abs/1201.5069} {arXiv:1201.5069}.
%%CITATION = 1201.5069;%%

\bibitem{Horowitz:2011wm}
W.~Horowitz, AIP
  Conf.Proc.\ 1441 (2012) 889--891,
\newblock \href {http://arxiv.org/abs/1108.5876} {arXiv:1108.5876}.
%%CITATION = 1108.5876;%%

\bibitem{Gubser:2008as}
S.~S. Gubser, D.~R. Gulotta, S.~S. Pufu, F.~D. Rocha, JHEP 0810 (2008) 052,
\newblock \href {http://arxiv.org/abs/0803.1470} {arXiv:0803.1470}.
%%CITATION = 0803.1470;%%

\bibitem{Chesler:2008uy}
P.~M. Chesler, K.~Jensen, A.~Karch, L.~G. Yaffe, Phys.Rev.\ D79
  (2009) 125015,
\newblock \href {http://arxiv.org/abs/0810.1985} {arXiv:0810.1985}.
%%CITATION = 0810.1985;%%

\bibitem{Ficnar:2012nu}
A.~Ficnar, Phys.\ Rev.\ D86 (2012) 046010, \href
  {http://arxiv.org/abs/1201.1780} {arXiv:1201.1780}.
%%CITATION = 1201.1780;%%

\bibitem{MoradHorowitz}
R.~Morad, W.~A. Horowitz, {\emph{in preparation}}.

\bibitem{Horowitz:2009pw}
W.~Horowitz, Y.~V. Kovchegov, Phys.Lett.\ B680 (2009) 56--61,
\newblock \href {http://arxiv.org/abs/0904.2536} {arXiv:0904.2536}.
%%CITATION = 0904.2536;%%

\bibitem{Dominguez:2011cy}
F.~Dominguez, D.~Kharzeev, E.~Levin, A.~Mueller, K.~Tuchin, Phys.Lett.\ B710 (2012) 182--187,
\newblock \href {http://arxiv.org/abs/1109.1250} {arXiv:1109.1250}.
%%CITATION = 1109.1250;%%

\bibitem{Chesler:2010bi}
P.~M. Chesler, L.~G. Yaffe, Phys.Rev.Lett.\ 106 (2011) 021601,
\newblock \href {http://arxiv.org/abs/1011.3562} {arXiv:1011.3562}.
%%CITATION = 1011.3562;%%

\bibitem{Chesler:2011nc}
P.~M. Chesler, Y.-Y. Ho, K.~Rajagopal, Phys.Rev.\ D85 (2012) 126006,
\newblock \href {http://arxiv.org/abs/1111.1691} {arXiv:1111.1691}.
%%CITATION = 1111.1691;%%

\bibitem{Andersen:2011sf}
J.~O. Andersen, L.~E. Leganger, M.~Strickland, N.~Su, JHEP 1108 (2011) 053,
\newblock \href {http://arxiv.org/abs/1103.2528} {arXiv:1103.2528}.
%%CITATION = 1103.2528;%%

\bibitem{El:2008yy}
A.~El, A.~Muronga, Z.~Xu, C.~Greiner, Phys.Rev.\ C79 (2009) 044914,
\newblock \href {http://arxiv.org/abs/0812.2762} {arXiv:0812.2762}.
%%CITATION = 0812.2762;%%

\bibitem{Wicks:2005gt}
S.~Wicks, W.~Horowitz, M.~Djordjevic, M.~Gyulassy, Nucl.Phys.\ A784 (2007) 426--442,
\newblock \href {http://arxiv.org/abs/nucl-th/0512076}
  {arXiv:nucl-th/0512076}.
%%CITATION = NUCL-TH/0512076;%%

\bibitem{Adare:2008cg}
A.~Adare, et~al., Phys.\ Rev.\ C77 (2008)
  064907,
\newblock \href {http://arxiv.org/abs/0801.1665} {arXiv:0801.1665}.
%%CITATION = 0801.1665;%%

\bibitem{Adare:2010de}
A.~Adare, et~al., Phys.Rev.\ C84
  (2011) 044905,
\newblock \href {http://arxiv.org/abs/1005.1627} {arXiv:1005.1627}.
%%CITATION = 1005.1627;%%

\bibitem{CMS:2012aa}
S.~Chatrchyan, et~al., Eur.Phys.J.\ C72 (2012)
  1945,
\newblock \href {http://arxiv.org/abs/1202.2554} {arXiv:1202.2554}.
%%CITATION = 1202.2554;%%

\bibitem{Adare:2010sp}
A.~Adare, et~al., Phys.\ Rev.\ Lett.\ 105 (2010)
  142301,
\newblock \href {http://arxiv.org/abs/1006.3740} {arXiv:1006.3740}.
%%CITATION = 1006.3740;%%

\bibitem{Horowitz:2011gd}
W.~Horowitz, M.~Gyulassy, 
  Nucl.Phys.\ A872 (2011) 265--285,
\newblock \href {http://arxiv.org/abs/1104.4958} {arXiv:1104.4958}.
%%CITATION = 1104.4958;%%

\bibitem{Horowitz:2011cv}
W.~Horowitz, M.~Gyulassy, J.Phys.\
  G38 (2011) 124114,
\newblock \href {http://arxiv.org/abs/1107.2136} {arXiv:1107.2136}.
%%CITATION = 1107.2136;%%

\bibitem{Chatrchyan:2012xq}
S.~Chatrchyan, et~al., 
  Phys.Rev.Lett.\ 109 (2012) 022301,
\newblock \href {http://arxiv.org/abs/1204.1850} {arXiv:1204.1850}.
%%CITATION = 1204.1850;%%

\bibitem{Uphoff:2012it}
J.~Uphoff, O.~Fochler, Z.~Xu, C.~Greiner, \href {http://arxiv.org/abs/1208.1970} {arXiv:1208.1970}.
%%CITATION = 1208.1970;%%

\bibitem{Buzzatti:2012pe}
A.~Buzzatti, M.~Gyulassy, \href {http://arxiv.org/abs/1207.6020}
  {arXiv:1207.6020}.
%%CITATION = 1207.6020;%%

\bibitem{Sharma:2012dy}
R.~Sharma, I.~Vitev, \href {http://arxiv.org/abs/1203.0329}
  {arXiv:1203.0329}.
%%CITATION = 1203.0329;%%

\bibitem{He:2012xz}
M.~He, R.~J. Fries, R.~Rapp, \href {http://arxiv.org/abs/1208.0256} {arXiv:1208.0256}.
%%CITATION = 1208.0256;%%

\bibitem{Cao:2012as}
S.~Cao, G.-Y. Qin, S.~A. Bass, \href {http://arxiv.org/abs/1209.5405}
  {arXiv:1209.5405}.
%%CITATION = 1209.5405;%%

\bibitem{Gossiaux:2012cv}
P.~B. Gossiaux, \href
  {http://arxiv.org/abs/1209.0844} {arXiv:1209.0844}.
%%CITATION = ARXIV:1209.0844;%%

\end{thebibliography}
\end{document}